\documentclass[fleqn,usenatbib,useAMS]{mnras}

\usepackage{newtxtext,newtxmath}

\usepackage{graphicx}	
\usepackage{amsmath}	
\usepackage{multicol}
\usepackage{longtable}
\usepackage{bm}	
\usepackage{pdflscape}

\let\VANthebibliography\thebibliography
\def\thebibliography{\DeclareRobustCommand{\VAN}[3]{##3}\VANthebibliography}
\usepackage[dvipsnames]{xcolor}
\usepackage[T1]{fontenc}
\usepackage{ae,aecompl}
\usepackage{newtxtext,newtxmath}
\usepackage[normalem]{ulem}
\usepackage{soul}
\usepackage{cancel}
\usepackage{adjustbox}
\usepackage{longtable}
\usepackage{tabularx}

\title[Jet power of Mkn 501]{Estimating the Jet Power from Broadband SED modeling of Mkn 501 for different particle distributions}

\author[H. Bora et al.]
{Hritwik Bora$^{1}$, \thanks{E-mail: hritwikbora@gmail.com}
	Rukaiya Khatoon$^{2}$, \thanks{E-mail: rukaiyakhatoon12@gmail.com}
	Ranjeev Misra$^{3}$, \thanks{E-mail: rmisra@iucaa.in}
	Rupjyoti Gogoi$^{1}$ \thanks{E-mail: rupjyotigogoi@gmail.com}
	\\
	$^{1}$Department of Physics, Tezpur University, Tezpur - 784028, Assam, India\\
	$^{2}$Centre for Space Research, North-West University, South Africa\\
	$^{3}$Inter-University Centre for Astronomy and Astrophysics, Post Bag 4, Ganeshkhind, Pune, Maharashtra - 411007 }

\date{Accepted 2024 March 4. Received 2024 March 1; in original form 2023 September 29}

\pubyear{2022}

\begin{document}
\label{firstpage}
\pagerange{\pageref{firstpage}--\pageref{lastpage}}
\maketitle

\begin{abstract}
	We consider the broadband spectral energy distribution of the high energy peaked (HBL) blazar Mkn 501 using \textit{Swift}-XRT/UVOT, NuSTAR and \textit{Fermi}-LAT observations taken between 2013 and 2022. The spectra were fitted with a one-zone leptonic model using synchrotron and synchrotron self-Compton emission from different particle energy distributions such as a broken power-law, log-parabola, as well as distributions expected when the diffusion or the acceleration time scale are energy dependent. The jet power estimated for a broken power-law distribution was $ \sim 10^{47} (10^{44})$ erg s$^{-1}$ for a minimum electron energy $\gamma_\text{min} \sim 10 (10^3)$. However, for electron energy distributions with intrinsic curvature (such as the log-parabola form), the jet power is significantly lower at {a} few times $ 10^{42}$ erg s$^{-1}$ which is a few percent of the Eddington luminosity of a $10^7$ M$_\odot$ black hole, suggesting that the jet may be powered by accretion processes. We discuss the implications of these results.
	
\end{abstract}




\section{INTRODUCTION}
Blazars are jetted objects, the subclass of Active Galactic Nuclei (AGN) where the jet is directly pointed towards the observer \citep{urry1995unified}. Significant properties include its Spectral Energy Distribution (SEDs) ranging from Radio - TeV gamma rays, non-thermal radiation, rapid variability, radio-loud features, and double hump structure \citep{urry1998multiwavelength,massaro2004log}. Blazars are broadly classified into BL Lacs and Flat Spectrum Radio Quasars (FSRQs) based on their optical emission lines. FSRQs show strong emission lines in the optical band whereas BL Lacs show faint emission lines \citep{1995PASP..107..803U,antonucci1993unified}. Blazars are also classified into three classes based on the peak of the emission of the synchrotron component. For low-Energy peaked (LBL)/Low - Synchrotron peaked (LSP) BL Lacs, the low-energy hump peaks at the optical/UV band. For High energy peaked (HBL)/high synchrotron peaked (HSP) BL Lacs, the component peaks in the X-ray band and the Intermediate peaked BL Lacs (IBL) have a peak in between the X-rays and optical/UV band \citep{ghisellini1997optical, 1998MNRAS.299..433F}. The broadband SED is composed of two peaks with the lower one attributed to synchrotron emission. The radiative processes responsible for the second component in the high-energy regime are still unresolved. In the leptonic model interpretation, the component is assumed to be due to inverse Compton (IC) scattering \citep{1992A&A...256L..27D, 1993ApJ...416..458D}. The seed photons for the IC processes may either be the synchrotron photons (i.e. the Synchrotron Self Compton (SSC) process) or they may be external photons from outside the jet \citep{1985ApJ...298..114M, 1993ApJ...416..458D, Khatoon_2022}.

One of the main challenges in understanding these systems, is the identification of the energy-generating process which results in these jets having such high powers, as inferred from broadband spectral fitting. For example, \cite{Paliya_2017} have fitted the spectra of several blazars (both $\gamma$-ray loud and quiet) and have inferred total jet powers with an average value of few times $ 10^{47}$ erg s$^{-1}$ and with some sources exceeding $10^{48}$ erg s$^{-1}$. Such jet powers exceed the Eddington luminosity of super-massive black holes and hence they cannot be powered by accretion and instead require alternate energy production processes (e.g., \cite{Ghisellini_2014}).  In this interpretation, most of the jet power is in the kinetic energy of the protons whose number density is assumed to be the same as that of the non-thermal electrons. The inclusion of electron-positron pairs may decrease the power requirement, but it is unlikely that the pair fraction is substantial due to Compton drag effects and unobserved spectral signatures \citep[e.g.][]{Sikora_2000, 10.1111/j.1365-2966.2007.12758.x, 10.1093/mnras/stw107}. If instead of a leptonic origin for the VHE, one considers a hadronic one, then the jet power required is even higher and can exceed $10^{49}$ erg s$^{-1}$ \citep[e.g.][]{10.1093/mnrasl/slv039, Abe_2023}.

Since the jet power is dominated by the number of protons, the inferred jet power depends on the number of relativistic electrons. Typically the energy distribution of electrons is assumed to be a broken power-law form and the number of electrons then depends on the lowest lorentz factor of the distribution i.e. $\gamma_\text{min}$, which is often assumed to be of the order of unity or tens. Employing large values of $\gamma_\text{min} \sim 1000$, as permitted by the data can lead to a reduction in the jet power. However, the rationale behind having the minimum energy of the electrons to be that high remains unclear. Even then, assuming $\gamma_\text{min} = 1000$, for the low flux state of Mkn 501, the jet power can still be as high as $\sim 10^{44}$ erg s$^{-1}$ \citep[e.g.][]{Abe_2023}.

The jet power estimations above have been primarily based on the assumption that the underlying electron energy distribution is a broken power-law (or a smoothened broken power-law). There have been some evidences based on the X-ray and $\gamma$-ray data that the observed spectrum has significant curvature deviating from a power-law \citep{massaro2004log, Tanihata_2004, tramacere}. This may indicate that the underlying electron energy distribution has a curvature which may have the shape of a log-parabola distribution \citep{massaro2004log}. Physically motivated models have also been used such as distributions where the high energy cutoff is due to radiative losses ($\gamma$-max models), or where the curvature is due to energy-dependent acceleration (EDA) or energy-dependent diffusion (EDD) \citep{Sinha_2017,10.1093/mnras/sty2003,Hota_2021,Khatoon_2022}. Typically, these different interpretations are often spectrally degenerate, but perhaps may be distinguished by studying how physical are the observed correlations between the best fit spectral parameters for different models \citep{Hota_2021,Khatoon_2022}. 

Electron energy distributions which are different from a broken power-law one, are expected to provide different estimates of the jet power. The motivation of this work is to estimate these different jet powers with statistical errors, by fitting a set of broadband SED of a bright blazar, with models having different electron distributions. To do this,  we consider the HBL source Mkn 501 at a redshift of z = 0.034 which is among the most well studied blazars with  VHE emission. It shows extreme HBL characteristics during flaring as reported by \cite{refId0} and \cite{Sahu_2020}. It was first detected in 1995 by the Whipple observatory \citep{1996ApJ...456L..83Q}. It is one of the brightest blazar sources in the catalog after Mkn 421, and has been studied through several multi-wavelength missions. Moreover, Mkn 501 shows significant curvature in the SED as investigated by \cite{Tavecchio_2001}, which may be described by a log parabola distribution \citep{2004A&A...422..103M}. As reported by \cite{Abe_2023}, during 15 years of swift observations, the \textit{Swift}-XRT flux rate was high from 2012 to 2014, with the highest X-ray flaring activity during March-October 2014. A recent multi-wavelength study by \cite{Furniss_2015} with NuSTAR observations reported hard X-ray variability in the timescale $\sim$ 7 hours in the range from 7-30 keV which was not previously observed. For this work, we analyzed multi-wavelength observations from \textit{Swift}-XRT/UVOT, NuSTAR, and \textit{Fermi}-LAT during the period from 2013 to 2022 covering the high and historically low activity state of Mkn 501, giving ten near simultaneous broadband SED, which we fit using different models. The section \ref{obs_data} describes the observations and the data analysis technique. In section \ref{broadband_model}, the broadband spectral analysis is reported, whose results are presented in section \ref{results}. We summarise and discuss the results in the section \ref{sec:summary}. In this work, the cosmological constant is considered in accordance with the $\Lambda$CDM model with $H_0$ = 71 km s$^{-1}$ Mpc $^{-1}$, $\Omega_\Lambda = 0.73$, $\Omega_M = 0.27$ \citep{Komatsu2011}.

\section{Observations and Data Analysis}\label{obs_data}

\subsection{NuSTAR}

Nuclear Spectroscopic Telescope Array (NuSTAR) is a space-based X-ray telescope used for nuclear spectroscopy. It was launched by NASA on 13 June 2012 which uses a Wolter Telescope. It consists of two Focal Plane modules FPMA and FPMB \citep{Harrison_2013}. Several observations of Mkn 501 were obtained between 2013 to 2022 which are summarized in Table \ref{tab:table1}.

The software package, "nupipeline" which is integrated with Heasoft-6.28, has been used for the generation of cleaned data files for the analysis. "XSELECT  V2.4k" was used for plotting the event fits files \textit{(cl.evt)} in ds9. Source extraction of 30 arcseconds and background of 60 arcseconds had been carried out for all the NuSTAR observations. The "\textit{nuproducts}" tool was used for the generation of spectrum files (PHA), ancillary files (ARF), and response matrix files (RMF). The \textit{"grppha"} tool was used to obtain the grouped spectra with all the obtained products to have counts of 30 per bin. 

\subsection{\textit{Swift}-XRT}

We considered observations undertaken by \textit{Swift}-XRT which are nearly simultaneous with the NuSTAR ones.  The detector has an area of 135 cm$^2$ covering an energy range of (0.2 - 10 keV). Observations for the instruments NuSTAR and \textit{Swift}-XRT are provided by NASA's HEASARC archive\footnote{\url{https://heasarc.gsfc.nasa.gov/}}. Corresponding to the NuSTAR observation during March 2022, there are two \textit{Swift}-XRT observations for which the analysis was carried out separately i.e. without summing the two \textit{Swift}-XRT spectra. The observation details are listed in Table \ref{tab:table1}. 

We have used the software package "xrtpipeline" which is integrated with Heasoft-6.28, for the generation of cleaned data and used "XSELECT V2.4K" for plotting the event fits files \textit{(cl.evt)} in ds9. For the observations taken during  2013, 2021, and 2022 the \textit{Swift}-XRT counts were comparatively higher than the ones during 2017 and 2018. Among the 10 observations, 9 observations of \textit{Swift}-XRT are in the Windowed Timing (WT) mode, which are free from pile-up. For all the non-piled-up observations, a source region of 10 pixels and a background region of 20 pixels are extracted. However, "Obs ID 00035023170" was found to be piled up since it is in the  "photon counting" (PC) mode. For this observation, we extracted an annular region with an inner radius of 7 pixels and an outer radius of 18 pixels, and a background region of 36 pixels  which was free from source contamination. Using "XSELECT  V2.4k", the light curve and spectrum are extracted, and using \textit{"xrtmkarf"} and \textit{"quzcif"} tools the ancillary (ARF) and response matrix files (RMF) were generated respectively. Like NuSTAR, \textit{Swift}-XRT spectrum was grouped such that there were at least 30 counts per bin.

\subsection{\textit{Swift}-UVOT}

The UV observations for the data sets were taken from \textit{Swift}-UVOT. The \textit{Swift}-UVOT covers both the optical and UV bands of the EM spectrum and consists of three optical filters (B, V, U) and three UV filters (uvw1, uvw2, uvm2) \citep{2005SSRv..120...95R}. However, for the observations under consideration, only the UV filters were available. The source and background regions of 8 arcsecs and 16 arcsecs were extracted. The tool "UVOTSOURCE" was used to calculate the magnitude in the AB magnitude system. We have used the relation $\frac{A_v}{E(B-V)} = 3.1$ for the magnitude correction, where A$_v$ is the galactic extinction and $E(B-V)$ is the interstellar reddening which has been taken to be 0.017 mag \citep{schlafly2011measuring}.

From the magnitude values, the flux was calculated with the help of photometric zero points and conversion factors adopted from \citet{2011AIPC.1358..373B, 10.1093/mnras/stw1516}. The flux points and energies of UVOT observations were converted to "XSPEC version 12.11.1" \citep{1996ASPC..101...17A} readable (PHA) format using the tool "\textit{ftflx2xsp}".

\subsection{\textit{Fermi}-LAT}

The Large Area Telescope onboard the Fermi Gamma-ray Space Telescope \textit{Fermi}-LAT, operates in the energy range from 20 MeV to more than 300 GeV \citep{Atwood_2009}. In order to get a significant detection and spectra, we considered the 4 months \textit{Fermi}-LAT observations keeping the NuSTAR observations in the middle of the total duration. We analyzed the data spanning from February 2013 to May 2022 using the open-source tools \textit{fermipy} \footnote{\url{https://fermipy.readthedocs.io/en/latest}}. We considered the region of interest (ROI) to be 15$^{\circ}$ to carry out the analysis and followed the standard procedure as mentioned in the documentation \citep{2017ICRC...35..824W}. The energy range considered was from 100 MeV to 300 GeV with evclass=128 and evtype=3. The latest instrument function (IRF) "P8R3 SOURCE V3" has been used. To generate XML files, the galactic diffusion "\textit{gll iem v07.fits}" and the isotropic background model "\textit{iso P8R3 SOURCE V3 v1.txt}" have been used. Using the latest \textit{Fermi}-LAT 4FGL catalog \citep{2020ApJS..247...33A}, we have acquired the XML model file encompassing all sources within the ROI. Along with it, the fourth Fermi source catalog providing the spectral models and parameters for the sources located within the ROI are considered. Finally, the flux points and energies of \textit{Fermi}-LAT observations are converted to XSPEC readable (PHA) format using the tool "\textit{ftflx2xsp}".

\begin{table*}
	
	\caption {\label{tab:table1} Summary of NuSTAR and \textit{\textit{Swift}}-XRT/UVOT observations of the source Mkn 501.}
	\centering
	\begin{adjustbox}{width = \textwidth, center}
		
		\begin{tabular}{lcccccccc}
			
			\hline
			\hline
			\multicolumn{5}{c}{NuSTAR} & \multicolumn{3}{c}{\textit{Swift}-XRT / UVOT} \\ 
			\hline
			&  Obs ID   & Date and Time  & MJD & Exposure (ks)                                      
			&  Obs ID   & Date and Time  & Exposure (XRT) (ks) & Exposure (UVOT) (ks)\\
			\hline
			
			S$_1$ &  60002024002 & 2013-04-13 T02:31:07 & 56395 & 18.27 &   
			
			00080176001 & 2013-04-13 T01:20:59 & 9.65 & 9.46 \\       
			
			S$_2$ & 60002024006  & 2013-07-12 T21:31:07 & 56485 & 10.85 &
			
			00030793233 & 2013-07-12 T00:54:59 & 1.02 & 1.01 \\
			
			S$_3$ & 60002024008  & 2013-07-13 T20:16:07 & 56486 & 10.34 & 
			
			00030793236 & 2013-07-13 T07:20:58 & 1.01 & 1.00 \\       
			
			S$_4$ &  60202049002 & 2017-04-27 T22:26:09 & 57870 & 23.15 &
			
			00035023156 & 2017-04-27 T23:02:56 & 1.07 & 1.06 \\
			
			S$_5$ &  60202049004 & 2017-05-24 T22:11:09 & 57897 & 23.83 &  
			
			00035023170 & 2017-05-24 T07:58:57 & 0.96 & 0.95 \\
			
			S$_6$ &  60466006002 & 2018-04-19 T15:06:09 & 58227 & 23.11 &                                 
			00035023212 & 2018-04-19 T22:19:57 & 1.00 & 0.98 \\
			
			S$_7$ &  60702062002 & 2021-07-13 T17:36:09 & 59408 & 20.63 &                                 
			00011184138 & 2021-07-12 T22:39:35 & 0.81 & 0.80 \\
			
			S$_8$ &  60701032002 & 2022-03-09 T10:01:09 & 59647 & 19.72 &                                 
			00011184179 & 2022-03-09 T02:58:36 & 0.99 & 0.97 \\
			
			S$_9$ &  60701032002 & 2022-03-09 T10:01:09 & 59647 & 19.72 &
			
			00011184180 & 2022-03-09 T22:19:57 & 1.02 & 1.01 \\
			
			S$_{10}$ &  60702062004 & 2022-03-27 T01:26:09 & 59665 & 20.29 &                                 
			00011184187 & 2022-03-27 T03:49:36 & 0.93 & 0.92 \\
			
			\hline
			\hline
			
		\end{tabular}
	\end{adjustbox}
\end{table*}

\section{Broadband Spectral Analysis}\label{broadband_model}
The source Mkn 501 is well known to have the X-ray emission from synchrotron radiation, however, the hump in the higher energy is attributed mainly to the inverse Compton (IC) process. In our broadband spectral fitting of Mkn 501, we have considered the leptonic model with a single emitting region. We assume a spherical region having a radius R, filled with a tangled magnetic field, B, and isotropic electron distribution, n($\gamma$), moving with bulk lorentz factor $\Gamma$ at an angle $\theta$ with respect to the observer's direction.  The seed photons of the IC process are the synchrotron photons i.e. the synchrotron self-Compton (SSC) process. We fit the broadband using a local model in XSPEC as described below.  For the spectral fit, we have used the "\textit{Tbabs}" model \citep{Wilms_2000} to account for the Galactic absorption. The hydrogen column density for the X-ray observations, ${N_\text{H}} = 1.69 \times 10^{20} $ $\text{cm}^{-2}$ was considered and kept fixed, which was obtained in the LAB survey \citep{2005A&A...440..775K}. However, for the UV the $N_\text{H}$ value was fixed at 0 since the UV flux was dereddened initially.
For error calculation, we have adopted Markov Chain Monte Carlo (MCMC) method using the python-based code  XSPEC\_EMCEE\footnote{\url{https://github.com/jeremysanders/xspec_emcee} }. We have used 50 walkers with 5000 iterations and burnt the first 2000. The errors were calculated within a confidence range of 2$\sigma$. EMCEE is an extensible, pure-Python implementation of Goodman \& Weare's Affine Invariant Markov chain Monte Carlo (MCMC) Ensemble sampler.

Since the objective is to fit the broadband spectra with different electron energy distributions, $n (\gamma)$, it is convenient to use a convolution model in XSPEC such that it can be used along with a function that represents the electron distribution. In particular, we follow \citet{Hota_2021,Khatoon_2022} to represent the synchrotron flux at an energy $\epsilon$ to be, 

\begin{equation}\label{flux}
F_\text{syn} (\epsilon) = \frac{\delta^3 (1+z)}{d_L^2} V \mathbb{A} \int_{\xi_\text{min}}^{\xi_\text{max}} f \left( \frac{\epsilon}{\xi^2} \right) n(\xi) d\xi
\end{equation}

\noindent where $d_L$ is the luminosity distance and V is the volume of the emission region, $ \mathbb {A} = \frac{\sqrt{3} \pi e^3 B}{16 m_e c^2\sqrt{C}}$ and $f(x)$ is the synchrotron emissivity function. Instead of using the electron's lorentz factor $\gamma$, the electron energy distribution is represented by $n (\xi)$. The transformation is given by $\xi=\gamma \sqrt \mathbb{C}$, where $ \mathbb {C}=1.36 \times 10^{-11}\frac{\delta B}{1+z}$ keV with $\delta$ as the doppler factor, B as the magnetic field, and z being the redshift of the source. Note that $\xi$ represents $\gamma$ and $\xi^2$ has dimension of keV. Similarly, the synchrotron self-Comptonization flux is determined as a function of $ F_\text{syn} (\epsilon)$ and $n(\xi)$. The expressions used for the flux are given in \cite{Sahayanathan_2018} and references therein, with $\gamma$ transformed to $\xi$. Thus we have a local XSPEC convolution model, \textit{sscicon}. which can be convolved with any particle distribution using the XSPEC model form (\textit{sscicon*n}($\xi$)). The convolution model \textit{sscicon} has the following parameters, bulk lorentz factor $\Gamma$, redshift $z$, magnetic field $B$, viewing angle $i$, and size $R$. The other parameters including normalization $K$, arise from the specific particle distribution form as described below. The model has the provision to have as a parameter, the jet power $P_\text{j}$, instead of normalization, $K$. We describe the different particle distributions used in this work below.

\subsection{{Logparabola model}}
We consider the particle density to be described by the log parabola function, 

\begin{equation}\label{lp}
n({\xi})=K \left (\frac{\xi}{\xi_r} \right)^{- \alpha - \beta \text{log} \left(\frac{\xi}{\xi_r} \right)}
\end{equation}

\noindent Here, particle spectral index is $\alpha$ at the reference energy $\xi_r$, while  $\beta$ and $K$ are the spectral curvature parameter and the normalization. The reference  $\xi_r^2$ was fixed at 1 keV during the spectral fit and the other three parameters $\alpha, \beta $ and norm $K$ were kept free.

\subsection{Broken Power Law}

We considered a Broken power law (BPL) distribution of the particles to fit the spectrum. 

\begin{equation}
n(\xi)=
\begin{cases}
K (\xi/1\sqrt{{\text{keV}}})^{-p} & \text{for} \hspace{3pt} \xi < \xi_\text{break} \\
K \xi^{q-p}_\text{break}(\xi/1\sqrt{{\text{keV}}})^{-q} & \text{for} \hspace{3pt} \xi > \xi_\text{break}  
\end{cases}
\end{equation}

\noindent Where $\xi_\text{break}$ is the break energy, p is the electron spectral index for $\xi$ < $\xi_\text{break}$ and q is the {electron spectral index} for $\xi > \xi_\text{break}$ {and the transformation is given by $\xi=\gamma \sqrt \mathbb{C}$}. {We consider the broken power-law particle distribution to exist only between a minimum and maximum $\gamma$ i.e. $n (\xi)$ is non-zero for $\xi_\text{min} < \xi <\xi_\text{max}$, where  $\xi_\text{min} = \gamma_\text{min} \sqrt\mathbb{C}$ and $\xi_\text{max} = \gamma_\text{max} \sqrt \mathbb{C}$}.

\subsection{Particle distribution with a maximum electron energy}
We consider the case where there is acceleration and electrons lose energy through the radiative processes. The evolution of the particles in the acceleration region is governed by the kinetic equation which is given by, 

\begin{equation}
\label{gamma}
\scriptsize \frac{\partial}{\partial \gamma} \left[ \left (\frac{\gamma}{\tau_\text{acc}}-\beta_s \gamma^2 \right)n_a \right ]+ \frac{n_a}{\tau_\text{esc}}=Q\delta(\gamma-\gamma_0)
\end{equation}
\noindent
Where the $\tau_\text{acc}$ and $\tau_{esc}$ are the acceleration and escape time scales of the electrons. $\beta_s$ is the cooling term is given by, $\beta_s=\frac{4}{3} \frac{\sigma_T B^2}{8 \pi m_e c^2}$,
where B is the magnetic field, $\sigma_T$ is the Thomson's cross-section, and $m_e$ is the mass of the electron. Q is the mono-energetic injection of electrons at minimum energy $\gamma_0$. The solution is given by, 

\begin{equation}
\scriptsize n(\xi)=K \xi^{-p}(1-(\xi/\xi_\text{max}))^{(p-2)}
\end{equation}    
where,
$p = \frac{\tau_\text{acc}}{\tau_\text{esc}} + 1$, $\xi_\text{max} = \gamma_\text{max}\sqrt\mathbb{C}$, $\gamma_\text{max} = \frac{1}{\beta_s\tau_\text{acc}}$ \\

\noindent{The particle distribution exists only for $\xi > \xi_\text{min}$, where $\xi_\text{min} = \gamma_\text{min} \sqrt \mathbb{C}$ }.

\subsection{Energy dependent time-scale models:}

We also employed models incorporating energy-dependent escape or acceleration timescales to provide a possible interpretation of the curvature in the observed spectrum \citep{Hota_2021,Khatoon_2022}.

\subsubsection{Energy dependent diffusion model (EDD)}

The diffusion takes place within a region containing a tangled magnetic field, causing the escape time scale to depend on the electron's gyration radius. In this context, the escape time scale ($\tau_\text{esc}$) maybe energy dependent as

\begin{equation}
\tau_{esc}=\tau_{esc,R} \left(\frac{\gamma}{\gamma_R} \right)^{-k}
\end{equation}

\noindent where, $\tau_{esc,R}$ corresponds to $\tau_{esc}$, when the electron energy is $\gamma_{R}mc^2$, and $k$ is the index of the assumed power-law  energy dependence. As $\gamma_R$ cannot be greater than the free streaming value $\tau_{esc,R}$ which restricts the equation to $\gamma < \gamma_R$, 
where $\gamma_R$ denotes the energy at which this limit is obtained.
If we make the assumption that $\gamma_R$ is significantly greater than any $\gamma$ of interest and ignore the synchrotron losses, we obtain the following electron energy distribution,

\begin{equation}
n(\xi)=Q_0 \tau_{acc} \sqrt\mathbb{C} \xi^{-1} exp \left[ -\frac{n_R}{k} \left( \left(\frac{\xi}{\xi_{R}} \right)^k- \left (\frac{\xi_{0}}{\xi_{R}} \right)^k \right) \right]
\end{equation}

\noindent where, $\xi_R=\sqrt\mathbb{C}\gamma_R$, $\xi_0=\sqrt\mathbb{C}\gamma_0$ and $\eta \equiv \frac{\tau_{acc}}{\tau_{esc,R}}$ \\

\noindent It is convenient to recast the distribution as follows,

\begin{equation}
n(\xi)=K(\xi)^{-1}exp \left [-\frac{\psi}{k}\xi^{k} \right]
\end{equation}

\noindent with $K$, $\psi$, and $k$ representing the free parameters. It can be shown that,

$$\psi = \eta_R (\mathbb{C}\gamma^{2}_{R})^{-k/2}=\eta_R \xi_{R}^{-k}$$
and that the normalization is given by,

$$K = Q_0 \tau_{acc} \text {exp} \left [\frac{n_R}{k} \left(\frac{\xi_0}{\xi_{R}} \right)^k \right]$$

\subsubsection{Energy dependent acceleration (EDA) model}
We consider the prospect that acceleration time-scale is energy-dependent and we assume its functional form to be given by,

\begin{equation}
\tau_{acc}=\tau_{acc,R} \left(\frac{\gamma}{\gamma_R} \right)^k
\end{equation} 

\noindent Where, $\tau_{acc,R}$ corresponds to $\tau_{acc}$, when the electron energy is $\gamma_{R}mc^2$ and  $k$ is the index of the energy dependence. Similar to the case of the EDD model, the corresponding electron energy distribution is given by,

\begin{equation}
\scriptsize n(\xi)=Q_0 \tau_{acc} \sqrt\mathbb{C} \xi_{R}^{-k} \xi^{k-1} \text{exp} \left [-\frac{n_R}{k} \left ( \left(\frac{\xi}{\xi_{R}} \right)^k- \left (\frac{\xi_{0}}{\xi_{R}}  \right )^k \right) \right]
\end{equation}

\noindent which can be {recast} as,

\begin{equation}
\scriptsize n(\xi)=K(\xi)^{k-1} \text{exp} \left [-\frac{\psi}{k}\xi^{k} \right] 
\end{equation}
where,\\
$$\psi = \eta_R (\mathbb{C}\gamma^{2}_{R})^{(-k/2)} = \eta_R \xi_{R}^{-k}$$

\noindent K is the normalization which is given by,

$$K = Q_0 \tau_{acc} \xi_{R}^{-k} \text{exp} \left [\frac{n_R}{k} \left(\frac{\xi_0}{\xi_{R}} \right )^k \right]$$

\section{Results}\label{results}
We performed the fitting of the observations listed in Table \ref{tab:table1} using the various particle distributions discussed in the preceding section.
We fixed the bulk lorentz factor, $\Gamma = 15$, and the viewing angle, $i = 0$, and allowed for the other parameters to vary. Note that for synchrotron and synchrotron-self Compton processes considered here, the relevant parameter is the doppler factor $\delta = 1/(\Gamma(1-\beta \text{cos}i)$ which for $i = 0$, $\delta = 2 \Gamma$.

\begin{figure*}
	\includegraphics[scale=0.6, angle=270]{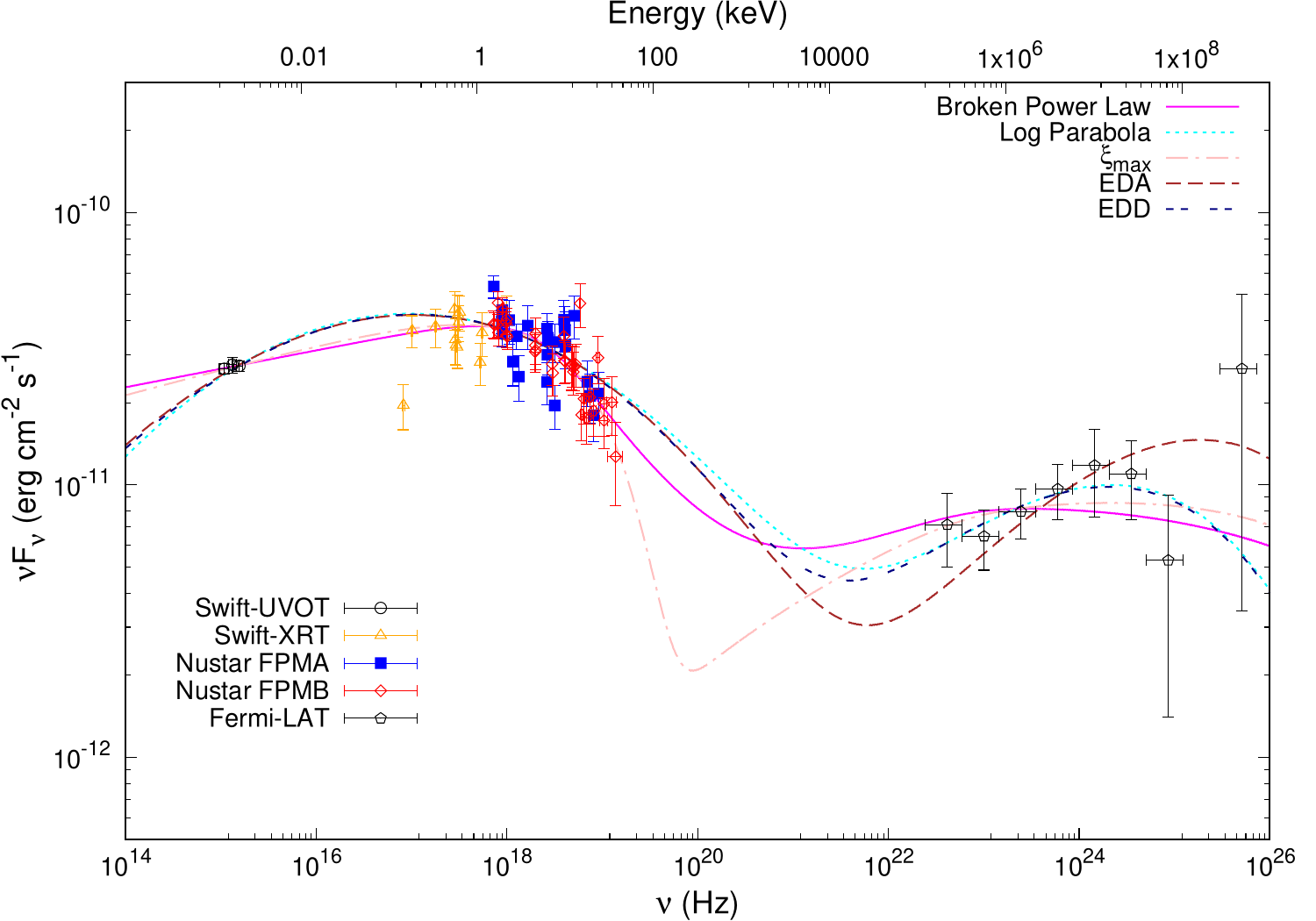}
	\caption{\centering Broadband SED plot of the observation s4 (MJD 57870) of Mkn 501 for all the five particle distribution models.} \label{sed_model}
\end{figure*}

For observations made during 2013 (listed as {S$_2$ and S$_3$} in Table \ref{tab:table1}), {none of the models considered provided a reasonable ﬁt}. This was because the \textit{Swift}-XRT spectra seemed to have a high energy excess, incompatible with the NuSTAR spectra. We were unable to find the reason for this and diferred a more detailed investigation of this to a later work. Thus, for these two observations, the \textit{Swift}-XRT spectra were omitted.
Using the different particle distributions, we achieved reasonable fits for all ten broadband spectra, and the corresponding best-fit parameters are listed in Table \ref{tab:bestfit}. A representative spectrum for the observation during MJD 57870 ( listed as S$_4$ in Table \ref{tab:table1}) is shown in Figure \ref{sed_model}, with the different best-fit model spectra overlayed. For some of the observations, especially the latter ones during 2022, the broken power-law distribution provides a better fit than the others. We note that the majority of the $\chi^2$ contribution originates from the X-ray spectra and in general for all the particle distribution considered, broadband SED is reasonably represented. 

\begin{figure*}
	\includegraphics[scale=1.2]{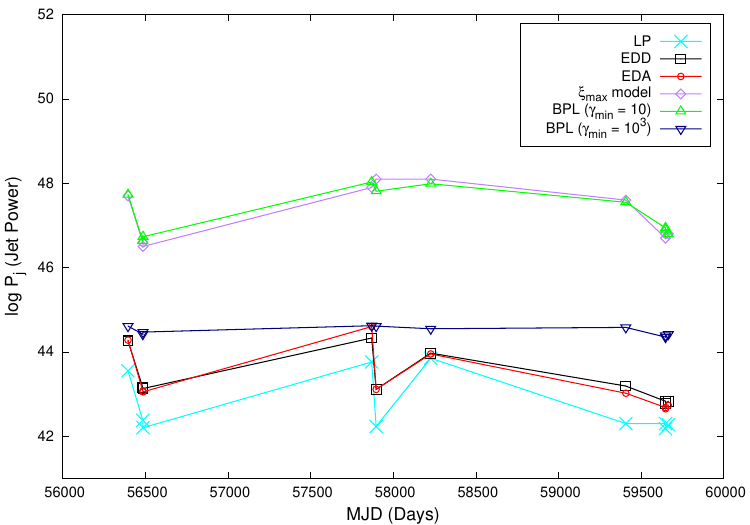}
	\caption{\centering {Variation of jet power for different particle distribution models.} \label{jp}}
\end{figure*}

\begin{figure*}
	\centering
	\includegraphics[scale=0.7]{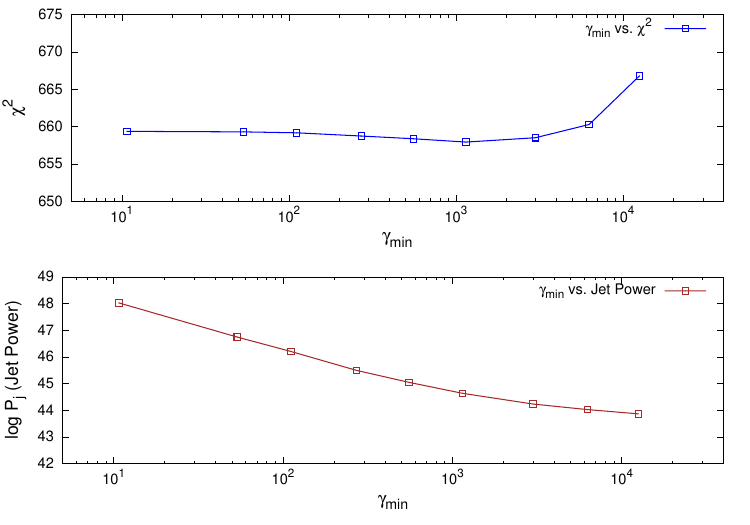}
	\caption{Variation of $\chi^2$ (top panel) and Jet Power (bottom panel) with $\gamma_\text{min}$ for the observation (MJD 57870) for BPL particle distribution.}
	\label{emin}
\end{figure*}

The estimated jet power differs significantly for the different particle distributions as shown in Figure \ref{jp} where the jet power is plotted against MJD. The models that have intrinsic curvature in the particle energy distributions (i.e. log-parabola, EDD and EDA) require nearly two orders of magnitude less jet power than the broken power-law and gamma-max distributions. For the latter two distributions, the jet power depends sensitively on the minimum energy of the electrons, $\gamma_\text{min}$. Figure \ref{emin} shows the variation of $\chi^2$ (top panel) and jet power as a function of $\gamma_\text{min}$ for the broken power-law distribution for a representative data set. As expected the jet power decreases with increasing $\gamma_\text{min}$ for large values, the $\chi^2$ increases because the UV flux is no longer consistent. For a value as large as $\gamma_\text{min} \sim 10^3$, the jet power for the broken power-law model is $\sim 10^{44}$ erg s$^{-1}$, which is still significantly larger than the power estimated for the log-parabola model in Figure \ref{jp}.

\begin{figure*}
	\centering
	\includegraphics[width=.5\textwidth]{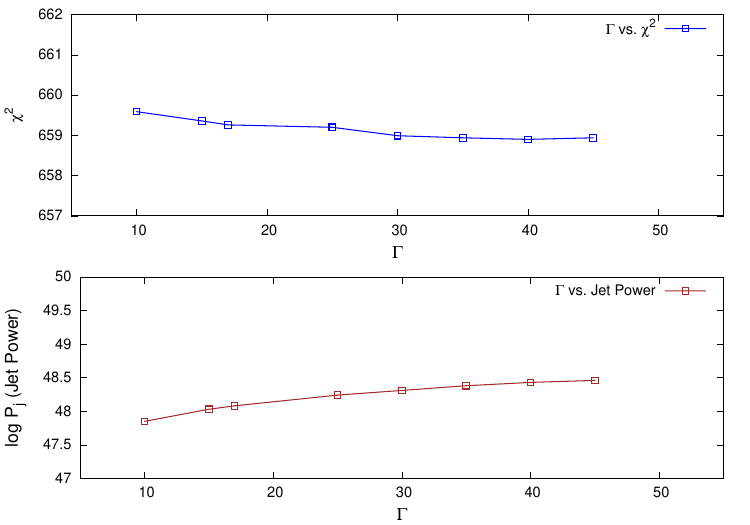}\hfill
	\includegraphics[width=.5\textwidth]{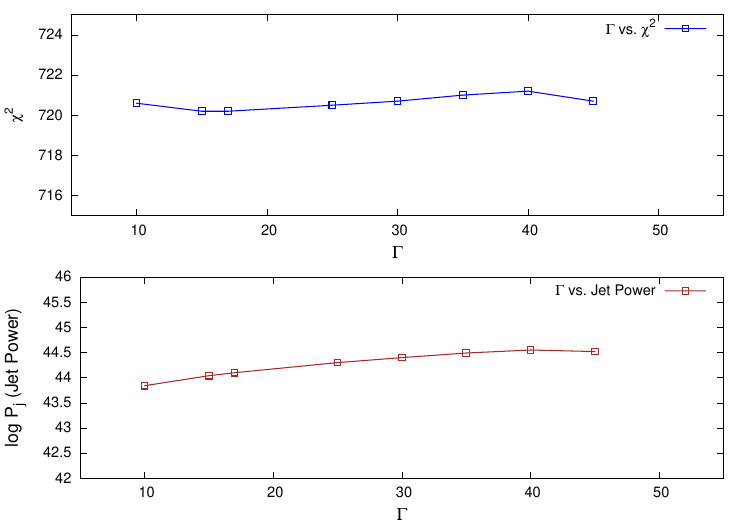}
	\caption{Variation of  $\chi^2$ (Top panel) and Jet Power (bottom panel) with $\Gamma$ for the observation (MJD 57870) for BPL (left panel) and LP (right panel) particle distribution.}
	\label{bulk_vs_jp}
\end{figure*}

In the above analysis, it was assumed that the bulk lorentz factor $\Gamma=15$. For a representative data set, the top panels of Figure \ref{bulk_vs_jp} show the variation of $\chi^2$ as a function of $\Gamma$ for the broken power-law (left panel) and log-parabola (right panel) models. The bottom panels show the dependence of jet power with $\Gamma$. As can be seen, $\Gamma$ is not constrained since the $\chi^2$ does not change with increasing $\Gamma$. The jet power increases with $\Gamma$ but only marginally, a factor of two increase when $\Gamma$ increases from 10 to 45. Thus the broad estimates of the jet powers presented here with $\Gamma = 15$ are not sensitive to that assumption.

\section{Summary and Conclusions}\label{sec:summary}

We collated ten broadband spectral energy distributions for Mkn 501, spanning 2013 to 2022, using data from Swift-UVOT, \textit{Swift}-XRT, NuSTAR, and \textit{Fermi}-LAT. We fitted the broad band spectra using a single zone leptonic model taking into account synchrotron and synchrotron-self Compton emission from a non-thermal electron distribution. Different electron energy distributions were considered such as a broken power-law, power-law with a maximum energy due to radiative cooling ($\gamma$-max model), log-parabola, energy-dependent diffusion time-scale (EDD) and energy-dependent acceleration time-scale (EDA). We determined that all these models adequately explained the broadband spectra and may be considered spectrally degenerate.

We estimated the total jet power with errors for the different observations and for different particle energy distributions. We found as reported before, that the jet power for broken power-law and $\gamma$-max models, was as high as $\sim 10^{47}$ erg s$^{-1}$ for a minimum lorentz factor $\gamma_\text{min} = 10$ and reduced to $\sim 10^{44}$ erg s$^{-1}$ for $\gamma_\text{min} = 10^3$, which was the maximum value allowed for spectral fitting. On the other hand, for the other energy particle distributions with intrinsic curvature, the jet power turned out to be $\sim 10^{43}$ erg s$^{-1}$, independent of $\gamma_\text{min}$. We show that these estimates for jet power are nearly independent of the chosen value of the bulk lorentz factor $\Gamma$.

\begin{figure*}
	\includegraphics[scale=0.8]{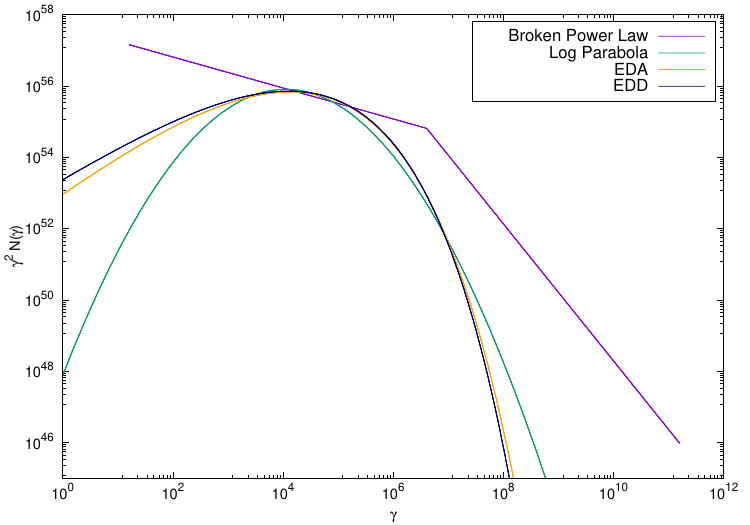}
	\caption{\centering {Particle energy distributions for the observation s3 (MJD 56486).} \label{pd}}
\end{figure*}

{To illustrate the reason for the difference in jet power among various particle distributions, we plotted Figure \ref{pd} that shows the particle distributions $\gamma^2 n(\gamma)$ as a function of $\gamma$ for the broken power-law, log-parabola, EDA, and EDD models for observation S$_3$. The peak of the function indicates the energies at which most of the particle energy dominates. In the case of the broken power-law distribution, the peak occurs at $\gamma_\text{min}$ and is significantly higher than for the other distributions, resulting in a higher jet power. It's noteworthy that the X-ray emission at $\sim$ 1 keV arises from electrons with $\gamma$ values of $\sim$ 1.6 × 10$^6$, 7.5 × 10$^4$, 1.25 × 10$^5$, and 1.25 × 10$^5$ for the broken power-law, log-parabola, EDA, and EDD models, respectively.}

The comparatively lower jet power estimated for the intrinsically curved particle energy distributions (especially the log-parabola one) of $\sim 10^{43}$ erg s$^{-1}$ makes it about a few percent of the Eddington luminosity for a $10^7$ M$_\odot$ blackhole, $L_\text{Edd} \sim 10^{45}$ erg s$^{-1}$. This has an important ramification that such a jet may be directly powered by the accretion process, which in turn may provide a framework to understand the jet-disc connections inferred from the correlation between disc and jet parameters in non-beamed systems \citep{10.1046/j.1365-8711.1999.02657.x,Ho_2001,Ghisellini_2014, 10.1093/mnras/stw1730}. Furthermore, the results will have a bearing on galaxy evolution studies which invoke AGN feedback through a powerful jet \citep{Fabian_2012, 10.1093/mnras/sty708}. A low jet power also indicates a smaller reservoir of energy in the jet, which may be depleted due to shocks and radiative losses in a shorter time scale. This provides an incentive to make a detailed study of the time evolution of such systems that have intrinsically curved particle distributions and hence lower jet power. 

The intrinsically curved particle distributions assumed in this work, are empirical in the sense that the curvature is due to an assumed energy dependence of the diffusion or acceleration time-scales. Clearly, a more detailed study considering the micro-physics of the processes needs to be undertaken to validate the assumed dependence. Finally, the analysis needs to be extended to different blazars, especially other kinds of blazars, such as FSRQs, to get a more comprehensive picture.

\section*{Acknowledgements}
We acknowledge the use of public data from the NuSTAR, \textit{Swift}-XRT/UVOT from NASA’s High Energy Astrophysics Science Archive Research Center (HEASARC), \textit{Fermi}-LAT data from Fermi Science Support Center (FSSC) of Goddard Space Flight Center. H.B. acknowledges various help received during the work from his colleagues Jyotishree Hota, Anshuman Borgohain, and Olag Pratim Bordoloi. H.B. also acknowledges Dr. Pranjupriya Goswami for the help received during the initial phase of the work. R.K. acknowledges the financial support of the NWU PDRF Fund NW.1G01487, the South African Research Chairs Initiative (grant No. 64789) of the Department of Science and Innovation and the National Research Foundation of South Africa. H.B. and R.G. would like to acknowledge IUCAA for their support and hospitality through their associateship program.

\section*{DATA AVAILABILITY}
The data and software used in this study can be accessed via NASA's HEASARC webpages and the Fermi Science Support Center (FSSC), for which the corresponding links and references are provided within the manuscript.

\begin{table*}

	\caption {\label{tab:bestfit} The table reports the best-fitting spectral parameters obtained by fitting the broadband spectrum with the one-zone leptonic model with various particle distributions. For all the models the bulk Lorentz factor is fixed at, $\Gamma = 15$. The magnetic field is in the units of Gauss (G), Radius of the emitting region \textit{R} and Jet Power $P_j$ are in logarithmic scale with the units of cm and erg s$^{-1}$. For the BPL distribution, $\xi_\text{max}=10^5$ $\sqrt{\text{keV}}$ and $\xi_\text{min} = 10^{-5}$ $\sqrt{\text{keV}}${, and for $\xi_\text{max}$ distribution,$\xi_\text{min} = 10^{-5} \sqrt{\text{keV}}$ have been considered.}}     
	
	\centering
	\begin{tabularx}{\textwidth}{ccccccccccc}
		
		\hline
		& & & & &  Log Parabola & & &   \\
		\hline
		
		& & Obs  & $\alpha$  & $\beta$ & B (G) & $ \text{log R} \text{(cm)}$  & log $P_{j}$  & $\chi^{2}_\text{red}(\text {dof})$\\
		\hline
		
		& &S$_1$ & 3.09$^{+0.01}_{-0.01}$  & 0.32$^{+0.01}_{-0.01}$ & 0.043$^{+0.01}_{-0.01}$ & 15.9$^{+0.01}_{-0.08}$ & 43.55$^{+0.02}_{-0.02}$ & 1.06 ({847})  \\   
		
		& &S$_2$ & 2.77$^{+0.01}_{-0.01}$ & 0.45$^{+0.02}_{-0.02}$ & 0.68$^{+0.30}_{-0.04}$ & 15$^{+0.18}_{-0.18}$ &  42.38$^{+0.002}_{-0.002}$&  1.04 ({680})  \\
		
		& &S$_3$ & 2.81$^{+0.01}_{-0.01}$ & 0.48$^{+0.02}_{-0.02}$ & 3.16$^{+1.07}_{-0.2}$ & 14.44$^{+0.08}_{-0.08}$ &  42.20$^{+0.001}_{-0.001}$ & 1.04({634})  \\
		
		& &S$_4$ & 3.08$^{+0.01}_{-0.01}$ &  0.25$^{+0.02}_{-0.02}$ & {0.39$^{+0.27}_{-0.29}$} & {15.26$^{+0.96}_{-1.2}$} &  {44.03$^{+0.003}_{-0.003}$} &  {1.19 (605)}\\
		
		& &S$_5$ & 3.81$^{+0.04}_{-0.03}$ & 0.79$^{+0.06}_{-0.05}$ & 0.048$^{+0.02}_{-0.004}$ & 16.03$^{+0.007}_{-0.14}$ &  42.23$^{+0.1}_{-0.06}$ &  {1.04} ({294})\\
		
		& &S$_6$ & 3.17$^{+0.04}_{-0.04}$ & 0.24$^{+0.01}_{-0.01}$ & 0.0063$^{+0.07}_{-0.006}$ & 16.83$^{+0.66}_{-2.04}$ & 43.85$^{+0.6}_{-0.47}$ &  0.88 ({201})\\
		
		& &S$_7$ & 3.34$^{+0.01}_{-0.01}$  & 0.62$^{+0.01}_{-0.01}$ & 1.52$^{+2.03}_{-0.77}$ & 14.67$^{+0.19}_{-0.32}$ & 42.30$^{+0.06}_{-0.07}$ &  1.18 (540)\\
		
		& &S$_8$ & 3.04$^{+0.01}_{-0.01}$  & 0.61$^{+0.02}_{-0.02}$ & 1.19$^{+1.2}_{-0.58}$ & 14.82$^{+0.4}_{-0.5}$ & 42.3$^{+0.06}_{-0.07}$ &  1.68 (791)\\
		
		& &S$_9$ & 3.05$^{+0.01}_{-0.01}$  & 0.63$^{+0.02}_{-0.02}$ & 1.58$^{+1.16}_{-0.87}$ & 14.72$^{+0.44}_{-0.44}$ & 42.18$^{+0.16}_{-0.03}$ &  {1.55 (797)}\\
		
		& & S$_{10}$ & 2.86$^{+0.01}_{-0.01}$  & 0.5$^{+0.02}_{-0.02}$ & 1.55$^{+0.16}_{-0.28}$ & 14.77$^{+0.24}_{-0.53}$ & 42.27$^{+0.14}_{-0.03}$ & 1.57 (956) \\

		\hline

		& & & & & Broken power law  & & &  \\
		\hline
		& Obs  & p & q & $\xi_\text{break}$ $\sqrt{\text{(keV)}}$ & B (G) (10$^{-3}$) & $ \text{log R} \text{(cm)}$ & log $P_{j}$ & $\chi^{2}_\text{red}(\text {dof})$ \\
		\hline
		
		& S$_1$ & 2.76$^{+0.01}_{-0.01}$  & 3.64$^{+0.02}_{-0.05}$ &  1.5$^{+0.07}_{-0.07}$ & 1.8$^{+0.002}_{-0.0004}$ & 17.49$^{+0.57}_{-0.51}$ &  {47.7$^{+0.003}_{-0.003}$} & 1.06 (846)  \\       
		
		&S$_2$ & 2.41$^{+0.02}_{-0.03}$ &  3.72$^{+0.16}_{-0.15}$ & 2.44$^{+0.26}_{-0.26}$ & {1.1$^{+0.26}_{-0.02}$} & 17.5$^{+0.0008}_{-0.12}$ & {46.6$^{+0.002}_{-0.002}$} & 1.08 (679)  \\
		
		&S$_3$ & 2.43$^{+0.03}_{-0.02}$ & 3.88$^{+0.49}_{-0.24}$ & 2.52$^{+0.44}_{-0.29}$ & 1.0$^{+0.0008}_{-0.0008}$ & 17.49$^{+0.01}_{-0.17}$ & {46.75$^{+0.003}_{-0.003}$} & 1.008(633)  \\
		
		&S$_4$ & 2.86$^{+0.02}_{-0.02}$ & 3.99$^{+0.42}_{-0.36}$ & 2.58$^{+0.50}_{-0.54}$ & {3.7$^{+0.001}_{-0.002}$} & {17.32$^{+0.6}_{-0.8}$} &  {48.01$^{+0.05}_{-0.07}$} & {1.09 (604)}\\
		
		&S$_5$ & 2.81$^{+0.05}_{-0.05}$ & 4.57$^{+0.15}_{-0.12}$ & 0.98$^{+0.13}_{-0.11}$ & 2.02$^{+0.001}_{-0.0002}$ & 17.50$^{+0.01}_{-0.15}$ & {47.8$^{+0.03}_{-0.03}$} & 0.94 (293)\\
		
		&S$_6$ &  2.94$^{+0.03}_{-0.03}$  & 4.69$^{+0.16}_{-0.1}$ &  0.88$^{+0.08}_{-0.03}$ &1.21$^{+0.0005}_{-0.0005}$ & 17.86$^{+0.12}_{-0.25}$ & {48.01$^{+0.003}_{-0.003}$} & 0.93 (200)\\
		
		&S$_7$ &  2.69$^{+0.02}_{-0.03}$  & 4.23$^{+0.11}_{-0.12}$ & 1.46$^{+0.11}_{-0.13}$ & 2.40$^{+0.001}_{-0.0001}$ & 17.30$^{+0.004}_{-0.24}$ &  {47.61$^{+0.03}_{-0.003}$} & 1.06 (539)\\
		
		&S$_8$ &  2.51$^{+0.02}_{-0.02}$  & 4.37$^{+0.16}_{-0.14}$ & 1.95$^{+0.09}_{-0.1}$ & 4.64$^{+0.003}_{-0.0003}$ & 17.01$^{+0.003}_{-0.26}$ & {46.9$^{+0.003}_{-0.03}$} &  1.13 (790)\\
		
		&S$_9$ &  2.49$^{+0.02}_{-0.02}$  & 4.3$^{+0.36}_{-0.13}$ & 1.86$^{+0.26}_{-0.1}$ & 1.24$^{+0.001}_{-0.0001}$ & 17.5$^{+0.001}_{-0.45}$ & {46.9$^{+0.003}_{-0.03}$} & {1.06 (796)}\\
		
		&S$_{10}$ &  2.45$^{+0.01}_{-0.01}$  & 4.03$^{+0.11}_{-0.10}$  &  2.19$^{+0.11}_{-0.1}$ & 1.07$^{+0.0002}_{-0.0005}$ & 17.5$^{+0.06}_{-2.25}$ &  {46.7$^{+0.003}_{-0.003}$} & 1.07 (955)\\
		\hline

		& & & & &  $\xi_{\text {max}}$   & & & \\
		
		\hline
		& &Obs  & p  & $\xi_{\text {max}}$ $\sqrt{\text{(keV)}}$ & B (G) (10$^{-3}$) & $ \text{log R} \text{(cm)}$ & log $P_{j}$ & $\chi^{2}_\text{red}(\text {dof})$\\
		\hline

		& & S$_1$ & 2.76$^{+0.03}_{-0.03}$  & 6.09$^{+0.35}_{-0.35}$ & 2.1$^{+0.009}_{-0.001}$ & 17.5$^{+0.008}_{-0.22}$ &  {47.8$^{+0.003}_{-0.003}$} & 1.09 (847)  \\       
		
		& &S$_2$ & 2.41$^{+0.01}_{-0.01}$ &  6.61$^{+0.27}_{-0.23}$ & 1.0$^{+0.0005}_{-0.0005}$ & 17.5$^{+0.0008}_{-0.12}$ & {46.74$^{+0.003}_{-0.003}$}&  1.04 (680)  \\
		
		& &S$_3$ & 2.36$^{+0.03}_{-0.02}$ &  7.9$^{+0.47}_{-0.36}$ & 3.2$^{+0.002}_{-0.0003}$ & 16.99$^{+0.002}_{-0.32}$ & {46.5$^{+0.003}_{-0.003}$} & 1.01({634})  \\
		
		& &S$_4$ & 2.81$^{+0.01}_{-0.01}$ & 8.48$^{+0.5}_{-0.48}$ & {2.2$^{+0.001}_{-0.002}$} & {17.49$^{+0.66}_{-1.05}$} & {47.9$^{+0.003}_{-0.003}$} & {1.09 (605)}\\
		
		& &S$_5$ & 2.91$^{+0.02}_{-0.01}$ & 3.76$^{+0.1}_{-0.1}$ & 1.0$^{+0.0005}_{-0.0005}$ & 17.84$^{+0.01}_{-0.14}$ & {48.15$^{+0.003}_{-0.003}$} & {0.97(294)} \\
		
		& &S$_6$ &  2.98$^{+0.02}_{-0.02}$  & 3.07$^{+0.12}_{-0.17}$ & 1.0$^{+0.0004}_{-0.0004}$ & 17.86$^{+0.009}_{-0.12}$ & {48.19$^{+0.003}_{-0.003}$} & 0.87 ({201})\\
		
		& &S$_7$ & 2.69$^{+0.01}_{-0.01}$  & 4.46$^{+0.15}_{-0.14}$ & 1.53$^{+0.0011}_{-0.0012}$ & 17.50$^{+0.003}_{-0.35}$ & {47.63$^{+0.003}_{-0.003}$} & 1.15 (540)\\
		
		& &S$_8$ & 2.43$^{+0.02}_{-0.01}$  & 5.37$^{+0.12}_{-0.13}$ &  4.14$^{+0.002}_{-0.0002}$ & 17.01$^{+0.005}_{-0.22}$ &  {46.7$^{+0.003}_{-0.003}$} &  1.2 (791)\\
		
		& &S$_9$ & 2.49$^{+0.02}_{-0.02}$  & 4.46$^{+0.12}_{-0.17}$ & 18.3$^{+0.012}_{-0.001}$ & 16.5$^{+0.001}_{-0.3}$ & {46.8$^{+0.003}_{-0.003}$}  &  1.2 ({797})\\
		
		& &S$_{10}$ & 2.45$^{+0.01}_{-0.009}$  & 5.36$^{+0.12}_{-0.08}$ & 1.35$^{+0.0001}_{-0.0001}$ & 17.5$^{+0.15}_{-2.02}$ & {46.8$^{+0.003}_{-0.003}$}  &  1.17 (956)\\
		\hline

		& & & & & EDA  & & &   \\
		
		\hline
		& & Obs  & k  & $\psi$ & B (G) & $ \text{log R} \text{(cm)}$ & log $P_{j}$ & $\chi^{2}_\text{red}(\text {dof})$\\
		\hline
		
		&  &S$_1$ & 0.13$^{+0.006}_{-0.008}$  & 2.21$^{+0.01}_{-0.01}$ & 0.01$^{+0.004}_{-0.0003}$ & 16.49$^{+0.009}_{-0.11}$ & 44.28$^{+0.03}_{-0.12}$ &  1.05 (847)  \\       
		
		& &S$_2$ & 0.20$^{+0.01}_{-0.009}$ &  1.91$^{+0.02}_{-0.01}$ & 0.16$^{+0.01}_{-0.05}$ & 15.49$^{+0.004}_{-0.08}$ &  43.07$^{+0.01}_{-0.04}$&  1.04 (680)  \\
		
		& &S$_3$ & 0.21$^{+0.01}_{-0.009}$ &  1.96$^{+0.01}_{-0.01}$ & 0.15$^{+0.05}_{-0.01}$ & 15.49$^{+0.01}_{-0.08}$ &  43.05$^{+0.003}_{-0.003}$ & 1.01({634})  \\
		
		& &S$_4$ & 0.10$^{+0.008}_{-0.006}$ &  2.17$^{+0.016}_{-0.019}$ & {0.22$^{+0.26}_{-0.01}$} & {15.5$^{+0.003}_{-0.64}$} &  {45.06$^{+0.003}_{-0.003}$} &  {1.17 (605)}\\
		
		& &S$_5$ & 0.25$^{+0.02}_{-0.01}$ & 2.99$^{+0.06}_{-0.04}$ & 0.58$^{+0.3}_{-0.03}$ & 15.08$^{+0.008}_{-0.20}$ &  43.11$^{+0.006}_{-0.006}$ &  0.96 (294)\\
		
		& &S$_6$ &  0.24$^{+0.01}_{-0.01}$  & 3.17$^{+0.04}_{-0.04}$ & 0.006$^{+0.07}_{-0.006}$ & 16.83$^{+0.66}_{-2.04}$ &  43.95$^{+0.007}_{-0.007}$ &  0.88 ({201})\\
		
		& &S$_7$ & 0.23$^{+0.005}_{-0.005}$  & 2.51$^{+0.01}_{-0.01}$ & 1.01$^{+1.2}_{-0.5}$ & 14.83$^{+0.27}_{-0.15}$ &  43.02$^{+0.09}_{-0.07}$ &  1.12 (540)\\
		
		& &S$_8$ & 0.26$^{+0.008}_{-0.008}$  & 2.22$^{+0.01}_{-0.01}$ & 0.71$^{+0.58}_{-0.41}$ & 15.00$^{+0.01}_{-0.01}$ &  42.68$^{+0.22}_{-0.07}$ &  1.52 (791)\\
		
		& &S$_9$ & 0.26$^{+0.008}_{-0.008}$  & 2.23$^{+0.01}_{-0.01}$ & 0.71$^{+0.16}_{-0.29}$ & 14.99$^{+0.48}_{-0.59}$ &  42.65$^{+0.2}_{-0.07}$ &  1.4 (797)\\
		
		& &S$_{10}$ & 0.23$^{+0.007}_{-0.007}$  & 2.03$^{+0.007}_{-0.007}$ & 0.81$^{+0.24}_{-0.13}$ & 15.00$^{+0.01}_{-0.01}$ &  42.73$^{+0.18}_{-0.16}$ &  1.4 (956)\\
		
		\hline
		
		& & & & & EDD  & & &  \\
		
		\hline
		& & Obs  & k  & $\psi$ & B (G) & $ \text{log R} \text{(cm)}$ & log $P_{j}$ & $\chi^{2}_\text{red}(\text {dof})$\\
		\hline
		
		& &S$_1$ & 0.14$^{+0.007}_{-0.007}$  & 2.07$^{+0.01}_{-0.01}$ & 0.018$^{+0.009}_{-0.001}$ & 16.33$^{+0.85}_{-0.63}$ & 44.26$^{+0.33}_{-0.20}$ &  1.05 (847)  \\       
		
		& &S$_2$ & 0.22$^{+0.01}_{-0.01}$ &  1.7$^{+0.01}_{-0.01}$ & 0.16$^{+0.061}_{-0.018}$ & 15.5$^{+0.021}_{-0.08}$ &  43.15$^{+0.2}_{-0.2}$&  1.04 (680)  \\
		
		& &S$_3$ & 0.24$^{+0.01}_{-0.01}$ &  1.73$^{+0.01}_{-0.01}$ & 0.15$^{+0.05}_{-0.01}$ & 15.49$^{+0.01}_{-0.1}$ &  43.13$^{+0.21}_{-0.25}$ & 1.01({634})  \\
		
		& &S$_4$ & 0.11$^{+0.007}_{-0.007}$ &  2.07$^{+0.01}_{-0.01}$ & {0.22$^{+0.16}_{-0.18}$} & {15.5$^{+0.009}_{-0.08}$} &  {45.1$^{+0.003}_{-0.003}$} & {1.17 (605)}\\
		
		& &S$_5$ & 0.28$^{+0.01}_{-0.01}$ & 2.72$^{+0.02}_{-0.02}$ & 0.54$^{+1.0}_{-0.3}$ & 15.1$^{+0.2}_{-1.1}$ &  43.11$^{+0.006}_{-0.006}$ &  0.96 ({294})\\
		
		& &S$_6$ &  0.26$^{+0.01}_{-0.01}$  & 2.92$^{+0.03}_{-0.03}$ & 0.003$^{+0.005}_{-0.003}$ & 17.0$^{+0.53}_{-2.3}$ & 43.97$^{+0.41}_{-0.49}$ & 0.87 ({201})\\
		
		& &S$_7$ &  0.26$^{+0.01}_{-0.01}$  & 2.26$^{+0.01}_{-0.01}$ & 0.64$^{+0.24}_{-0.05}$ & 15.0$^{+0.01}_{-0.15}$ & 43.19$^{+0.09}_{-0.14}$ & 1.11 (540)\\
		
		& &S$_8$ &  0.29$^{+0.01}_{-0.01}$  & 1.95$^{+0.01}_{-0.007}$ & 0.54$^{+0.4}_{-0.2}$ & 15.09$^{+0.5}_{-0.6}$ & 42.83$^{+0.22}_{-0.16}$ & 1.5 (791)\\
		
		& &S$_9$ &  0.3$^{+0.01}_{-0.01}$  & 1.95$^{+0.01}_{-0.01}$ & 0.71$^{+0.37}_{-0.24}$ & 15.00$^{+0.53}_{-0.56}$ & 42.77$^{+0.18}_{-0.08}$ &  1.38 (797)\\
		
		& &S$_{10}$ & 0.26$^{+0.01}_{-0.009}$  & 1.78$^{+0.01}_{-0.01}$ & 0.81$^{+0.08}_{-0.13}$ & 15.00$^{+0.55}_{-0.65}$ & 42.83$^{+0.19}_{-0.10}$ & 1.42 (956)\\
		\hline

	\end{tabularx}
\end{table*}

\bibliographystyle{mnras}
\bibliography{example} 








\bsp	
\label{lastpage}
\end{document}